# O-Band Differential Phase-Shift Quantum Key Distribution in 52-Channel C/L-Band Loaded Passive Optical Network

**Bernhard Schrenk, Michael Hentschel, and Hannes Hübel**
*AIT Austrian Institute of Technology,*
*Center for Digital Safety&Security,*
*1210 Vienna, Austria.*
*Author e-mail address: bernhard.schrenk@ait.ac.at*

**Abstract:**
A cost-effective QKD transmitter is evaluated in a 16km reach, 2:16-split PON and yields $5.10^{-7}$ secure bits/pulse. Co-existence with 20 down- and 1 upstream channel is possible at low QBER degradation of 0.93% and 1.1%.

## I. Introduction

Quantum communication systems have made significant leaps towards commercialization during the past years. Yet, a remaining challenge is the practical network integration. The co-existence of classical and quantum channels is to be facilitated at marginal cost, which prevents the use of dark fibers as these would drive operating expenditures. Moreover, brown-field deployment dictates the classical channel count and their launched power. In such a zero-touch scenario the high power difference of >70 dB between classical and quantum signals requires a careful crosstalk management: Although the optical background of the source lasers is confined to the semiconductor material gain spectrum, stimulated Raman scattering at passive components and transmission fibers renders the primary impairment due to its wide spectral tails. These cause in-band crosstalk noise even though quantum and classic channels might be spaced by more than 200 nm [1]. Narrow filtering such as it is inherent to coherent reception in continuous-variable quantum communication can minimize these detrimental effects [2]; however, the high degree of system complexity does not bode well with the pressing cost requirements of short-reach applications. Alternative discrete-variable systems solve this problem by locating the quantum signal at the Raman dip in the classical wavelength spectrum while reducing the launched power of classical channels by >10 dB [3].

In this work we integrate a low-cost laser-based O-band differential phase-shift (DPS) transmitter [4] in a 16 km reach, 2:16 split passive optical network (PON). We evaluate the sensitivity to the classical channel count and show that co-existence with up to 52 unsuppressed channels leads to a QBER of 4.7% at a raw key rate of 2.7 kb/s.

## II. Differential Phase Shift QKD in Coexistence with Classical PON Signals

In contrary to legacy PON standards, recent access efforts such as NG-PON2 suggest to dedicate the O-band at 1310 nm for the quantum signals since classical signals, together with a possible WDM overlay for 5G fronthauling, will remain in either the C- or the L-band. In such a wideband WDM context the alignment of the quantum signal in the Raman dip next to a single or few classical carriers [3] would be hard to facilitate. Yet, O-band quantum channels require a thorough investigation given the higher transmission losses and brown-field deployments where splitting loss in a filterless optical distribution network (ODN) cannot be simply bypassed.

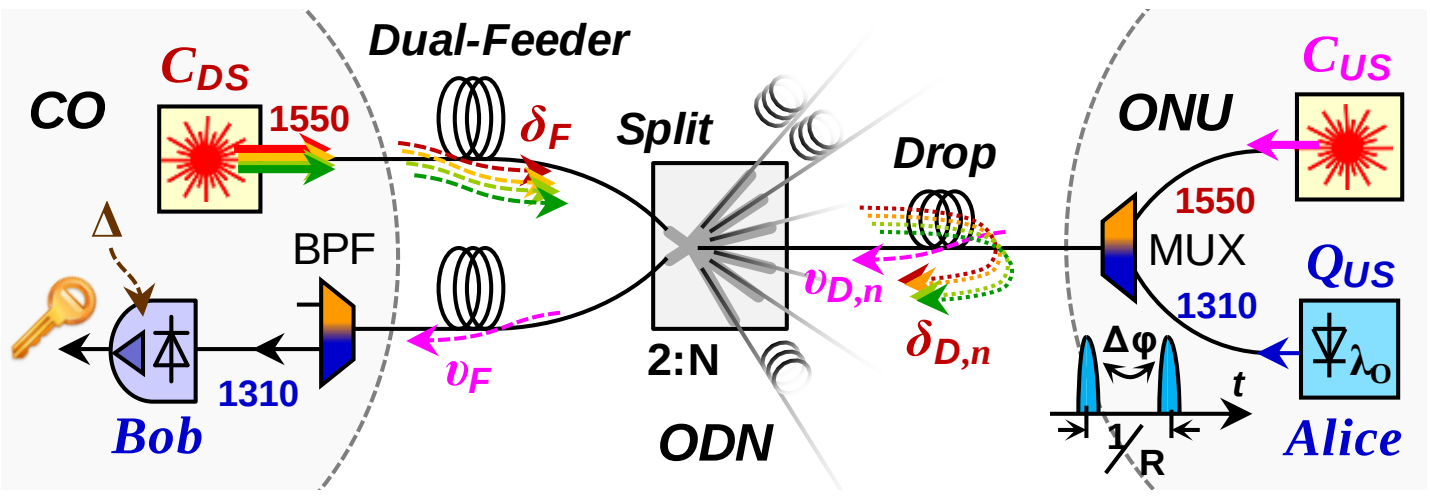

Fig. 1. Stimulated Raman scattering arising at the ODN of a PON.

The bidirectional nature of Raman scattering greatly determines the impact on the quantum channels and their integration approach. Classical upstream signals $C_{US}$ will generally feature less lanes (i.e., 4 wavelengths for NG-PON2) than the downstream $C_{DS}$ (WDM overlay) so that the quantum channel $Q_{US}$ is likely to be implemented in the upstream direction. A dual-feeder ODN with high-directivity 2:$N$ splitter therefore avoids Raman noise $\delta_F$ at the dominant feeder length, while noise $\delta_D$ arising at any of the $N$ drop fibers, that will eventually fall within the upstream direction, will have a much lower magnitude due to the shorter drop length and splitting loss. It shall be noted that although the downstream passes the splitter twice to reach the upstream receiver at Bob, the large number of $N$ drop fibers compensates for the double-pass assuming a uniform splitting loss at all ports. Moreover, since classical upstream transmission is subject to time division multiple access (TDMA), the overall noise contribution over all $N$ drop fibers corresponds to a single continuous-mode signal, that nevertheless traverses the PON over its entire reach of drop ($\upsilon_D$) and feeder ($\upsilon_F$) fibers. Although ODN loss attenuates Raman noise as it does the quantum signals, the constant dark counts Δ at Bob's SPAD receiver introduce a limit for the compatible ODN loss budget.

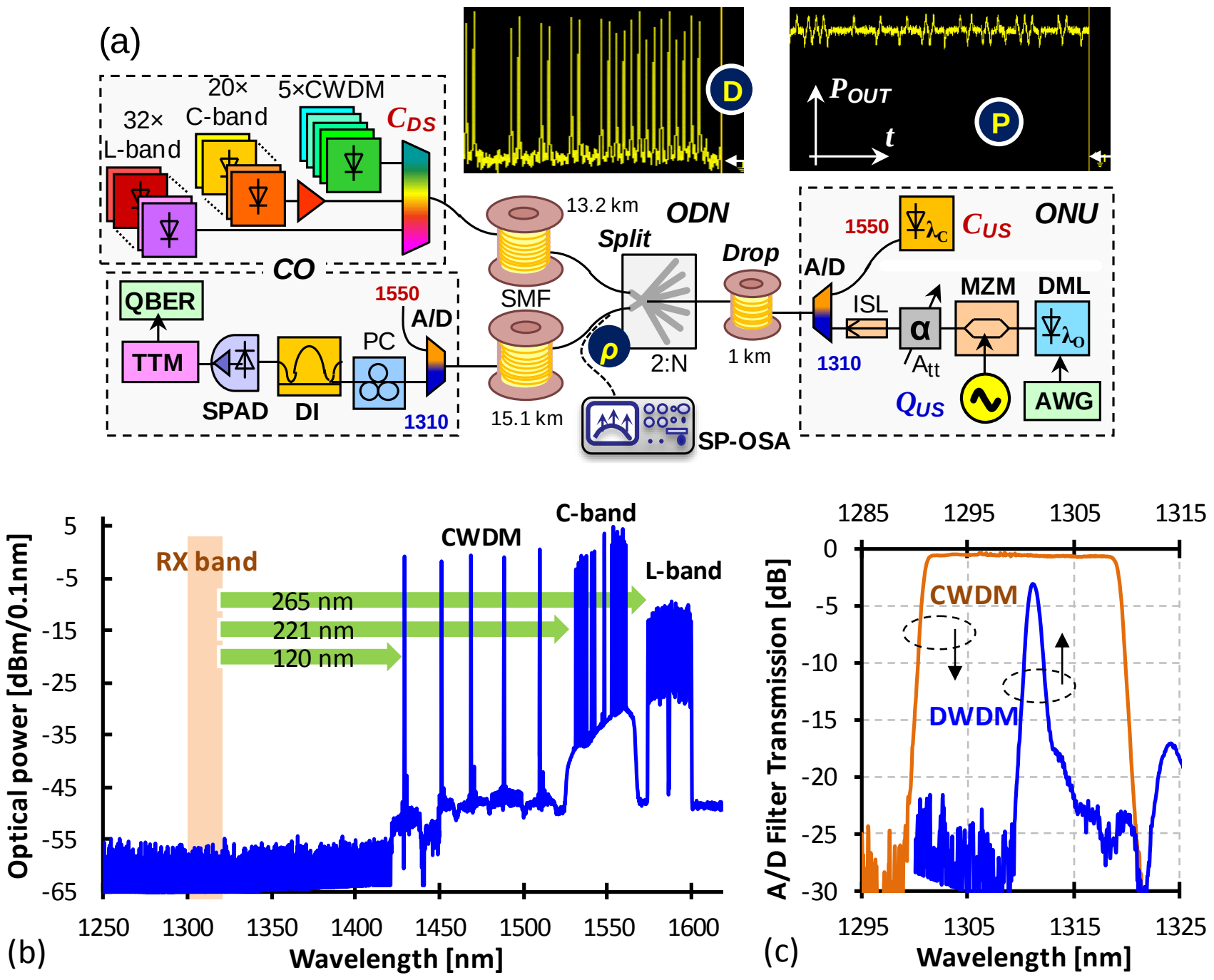

Fig. 2. (a) Experimental setup for DPS QKD in co-existence regime. (b) Downstream network load. (c) 1310-nm DWDM filter transmission.

## III. Experimental Setup and Network Load

Figure 2a presents the setup for the experimental evaluation of the DPS scheme in a PON scenario. The DPS transmitter is located at the optical network unit (ONU) and builds on a directly chirp-modulated laser (DML) at 1310 nm as phase modulator (inset P) and a pulse carver that cuts the symbol edges that appear in the pattern at DML output. Since neither a 1310 nm externally modulated laser (EML) nor an O-band electro-absorption modulator were available, a Mach-Zehnder modulator (MZM) was used for pulse carving. However, an EML-based DPS transmitter has been recently demonstrated at 1550 nm [4]. The transmitter is completed by an attenuator for setting a mean photon number of $\mu = 0.1$. An isolator prevents external probing.

The DPS receiver is hosted at the central office (CO) and includes a 1-ns delay interferometer (DI) adjusted to the DPS symbol rate of 1 Gbaud and a free-running InGaAs single-photon avalanche photodetector (SPAD) with a detection efficiency of 10%. The detector events are registered by a time-tagging module and processed in real-time to estimate the quantum bit error ratio (QBER) and raw key rate. A manual polarization controller (PC) was used to optimize the response of the DI yielding the demodulated carved eyes (inset D in Fig. 2a).

The co-existence of the quantum channel ($Q_{US}$) with classical channels has been evaluated by loading the network with (*i*) 32×100-GHz sliced self-seeded L-band channels, (*ii*) 20 C-band channels and (*iii*) 5 CWDM channels from 1430 to 1510 nm in downstream direction ($C_{DS}$) and a single or 20 upstream channel(s) at 1550 nm ($C_{US}$). The optical spectrum for the downstream channels is shown in Fig. 2b. The average per-channel power levels for L-, C-band and CWDM channels were -1.9, 2.5 and -0.7 dBm. Quantum and classical channels are multiplexed through 1310 nm add-drop (A/D) filters at ONU and CO, while the ODN was left filterless to conform to brown-field deployments. In order to achieve a good suppression of Raman noise, an O-band DWDM A/D filter was constructed for the quantum receiver. The transmission is shown in Fig. 2c and features a FWHM bandwidth of 1.22 nm, which improves the noise rejection with respect to CWDM filtering by 11.9 dB.

The ODN is composed of a 13.2 and 15.1 km dual-feeder fiber for down- and upstream, a 2:N splitting stage and a 1 km drop fiber. ITU-T G.652.B compatible standard single-mode fiber (SMF) has been used and the average loss experienced at 1310 nm was 0.37 dB/km. Together with the high splitting loss the 1310 nm allocation firstly seems unfavorable for the quantum signals. As we will prove, it trades well with reduced Raman noise.

## IV. Noise Contribution due to Stimulated Raman Scattering and Impact on DPS QKD Performance

For characterization purposes the back-to-back DPS QKD performance was evaluated according to a dark-fiber scenario serving as a reference. Fig. 3a shows the raw key rate as function of the optical loss budget. Even for a high 26 dB loss budget a rate of ~1 kb/s can be obtained. However, lower QBER values of ~3% are found at a budget of 15-19 dB since the dark counts Δ of the SPAD are standing in a more favorable, smaller ratio to the photon counts. At lower optical budgets the SPAD saturates and effects such as after-pulsing lead to a QBER increase.

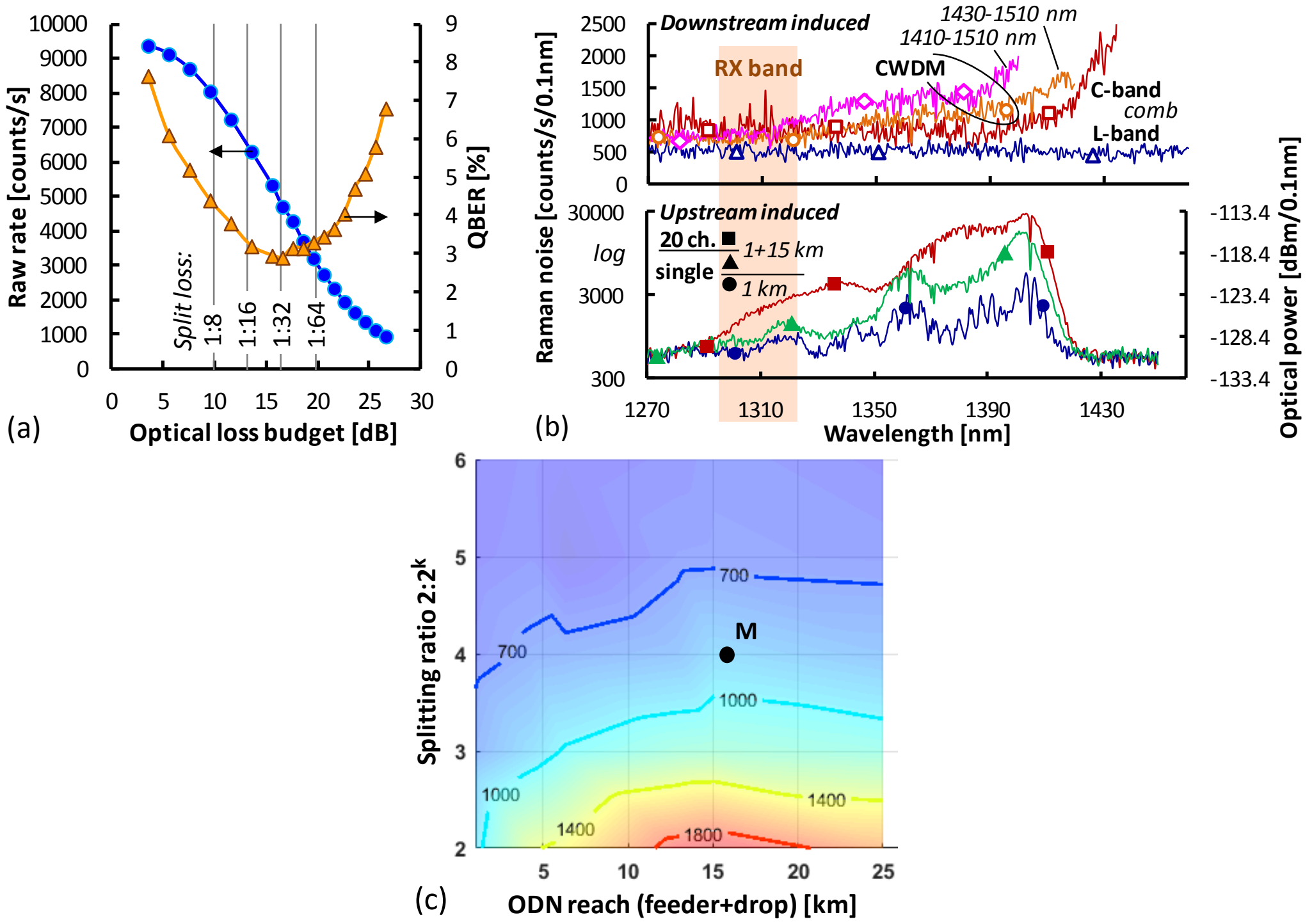

Fig. 3. (a) DPS QKD back-to-back performance. (b) Raman noise spectrum induced by classical down- and upstream channels. (c) Received Raman noise contribution at DPS QKD receiver as function of PON splitting ratio and ODN reach, including detector dark counts (Δ = 520 c/s).

Moreover, the Raman noise generated by various classical channels was characterized at the upstream-side output of the 2:*N* splitter (*ρ* in Fig. 2a) by means of single-photon optical spectral analysis (SP-OSA). To allow for a wider spectral acquisition range, the narrow A/D filter at Bob was replaced by two O/C-waveband splitters. Figure 3b presents the Raman noise spectrum for various combinations of classical channels and ODN reaches without extra splitting loss. The downstream-induced noise is pronounced for the closer CWDM channels (◇,○). Inclusion of a sixth channel at 1410 nm (◇) already shifts the first Raman shoulder close to the reception band. The far yet highly populated C-band comb (□) does not induce stronger noise than 5 CWDM channels (○), while the even farther L-band comb (△) does not show Raman contributions above the dark count level of the SP-OSA detector. The upstream Raman noise dominates the contribution that is induced by the downstream. Even a single channel transmitted over the intended ODN reach of 16 km (◆) shows a strong peak from 1355 nm onwards. However, the contribution at the reception band at 1310 nm already falls off strongly due to the seed wavelength in the C-band. Multi-channel upstream transmission, such as shown for 20 lanes (■), is nevertheless flooding the 1310 nm band.

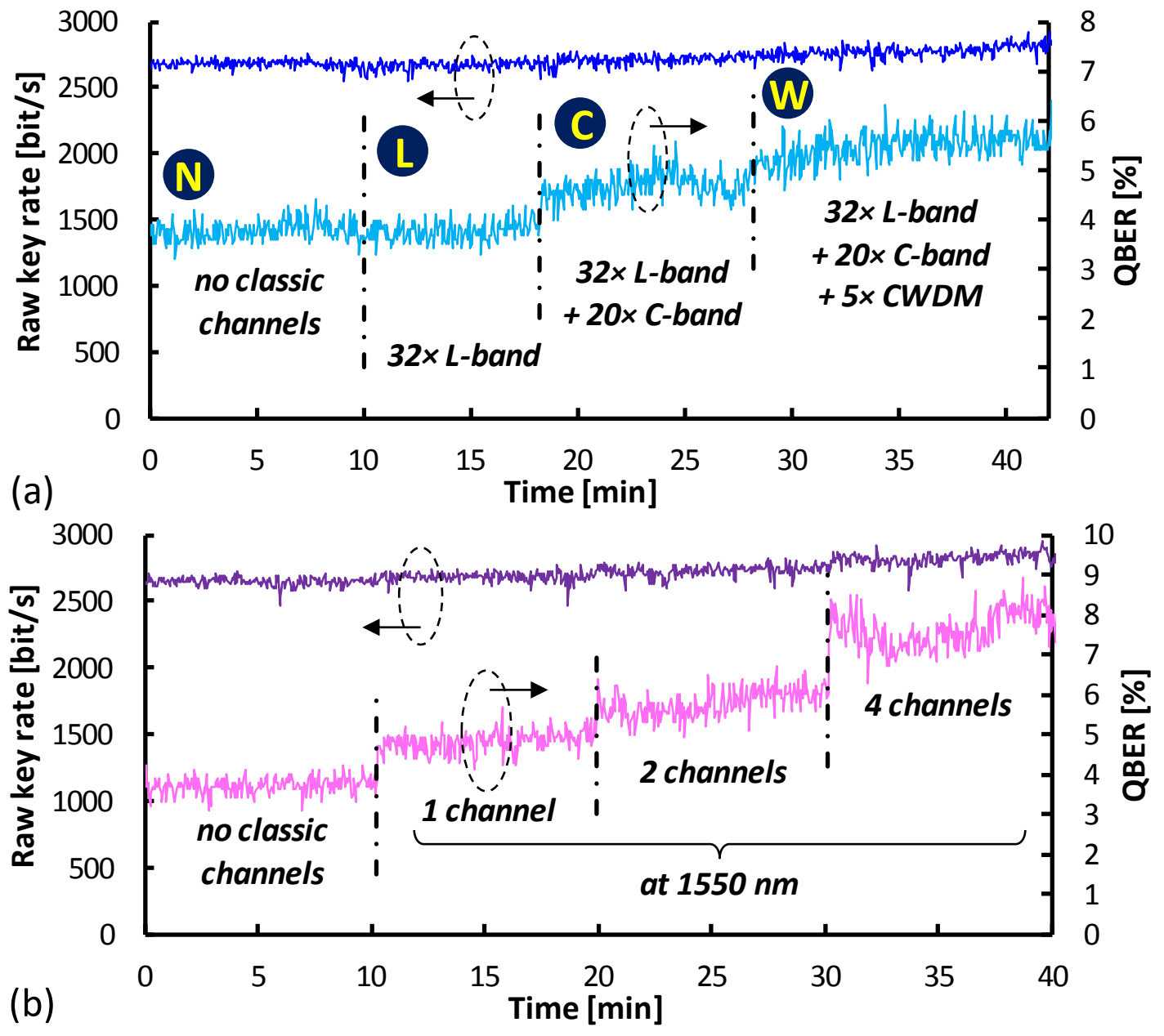

Fig. 4. Raw key rate and QBER for DPS-QKD in co-existence with various classical (a) down- and (b) upstream network load configurations.

The Raman noise has been further characterized in the PON configuration and the original QKD receiver that features the A/D filter, the DI and the SPAD, and applying the same settings for photon count registration as during the later QKD measurements. Figure 3c shows the received upstream Raman noise for various combinations of ODN reach and PON split 2:$N$, with $N = 2^k$, including ~520 dark counts/s of the SPAD receiver. Measurements have been taken for equivalent DWDM filtering, meaning that a CWDM A/D filter has been applied together with a single classical C-band channel reduced in its launched power by the ratio of passband widths between CWDM and DWDM filters. This is due to the unavailability of a DPS emitter whose wavelength would align with the 1301 nm center wavelength of the DWDM filter as shown in Fig. 2c. The concentrated splitting loss greatly determines the generated Raman noise as the classical pump for the feeder section and the generated noise at the drop fiber are both attenuated. Raman noise is then pronounced for longer fiber spans. A saturation effect can be noticed at ~15 km, which is explained by the extra loss that is introduced for longer transmission spans. A peak Raman contribution of ~360 c/s can be anticipated for a 2:16 split after subtracting the constant dark count rate.

Finally, the DPS QKD performance has been evaluated in terms of raw key rate and QBER for transmission over a 2:16 split, 16 km reach PON (point M in Fig. 3c) in presence of classical channels. Temporal filtering within 30% of the symbol period can be applied due to the carved quantum signal, therefore cutting part of the Raman noise. Figure 4a presents the performance for simultaneous transmission of classical downstream signals. Reference is made to the case without co-existence of classical channels (N), for which a raw key rate of 2.7 kb/s and a QBER of 3.77% are obtained. This intrinsic QBER is attributed to the low-cost DPS transmitter based on the chirp modulation principle [4]. It is estimated that ~500 bit/s of secure key remain after reconciliation [5]. There is no change for the far 32 L-band channels (L), while the QBER degrades by 0.93% when the C-band comb consisting of 20 channels is added (C). Further network loading with 5 CWDM channels from 1430 to 1510 nm leads to an additional degradation of 0.78% (W) due to the vicinity to the 1310 nm quantum channel.

The QKD performance under continuous-mode classical upstream transmission shows a QBER degradation of 1.1, 2.1 and 3.9% for 1, 2 and 4 C-band channels, respectively (Fig. 4b). This confirms that directional upstream transmission dominates the noise arising from a much larger number of classical downstream channels.

## V. Conclusions

A low-cost laser-based DPS QKD transmitter has been evaluated in a PON scenario under co-existence with classical channels in the S-/C-/L-bands. Raman noise at the O-band quantum channel is dominated by C-band upstream signals unless spectrally close CWDM downstream channels are lit. The QBER penalties arising from 20-channel downstream and a continuous single-channel upstream C-band channel(s) in a 16 km reach, 2:16 split PON are 0.93% and 1.1%, respectively. Yet, the QBER is below the 5% threshold for which a secure key can be obtained.

## VI. Acknowledgement

This work has received funding from the European Union's Horizon 2020 research and innovation programme under grant agreement No 820474.